\documentstyle[11pt,aaspp4]{article}
\def\kms                 {km\thinspace s$^{-1}$}

\received{24 November 1998}
\accepted{27 January 1999}

\lefthead{Lidman et al.}
\righthead{Spectroscopy of PKS~1830-211}

\begin{document}

\title{The redshift   of  the   gravitationally  lensed   radio source
PKS~1830$-$211\footnote{Based  on   observations  collected  at    the
European Southern Observatory, La Silla, Chile (ESO Program 61.B-0413)
} }

\author{C. Lidman}
\affil{European  Southern Observatory,   Casilla 19001, Santiago   19,
Chile; clidman@eso.org}

\author{F. Courbin}
\affil{Institut d'Astrophysique et  de G\'eophysique,  Universit\'e de
Li\`ege,     Avenue    de  Cointe  5,    B-4000    Li\`ege,   Belgium;
Frederic.Courbin@ulg.ac.be\\ URA  173  CNRS-DAEC, Observatoire   de
Paris, F-92195 Meudon Principal C\'edex, France}

\author{G. Meylan}
\affil{European  Southern Observatory,  Karl-Schwarzschild-Strasse  2,
D-85748 Garching bei M\"unchen, Germany; gmeylan@eso.org}

\author{T. Broadhurst, B. Frye and W.J.W. Welch}
\affil{Berkeley  Astronomy    Department,  University   of California,
Berkeley,   CA    94720,         U.S.A.;      tjb@wibble.berkeley.edu,
bfrye@worms.berkeley.edu, WWelch@astro.berkeley.edu}

\begin{abstract}

We report on  the  spectroscopic identification  and the long  awaited
redshift  measurement of the   heavily obscured, gravitationally lensed
radio source PKS~1830-211,  which was   first   observed as a  radio
Einstein  ring.   The NE   component  of  the   doubly imaged  core is
identified, in our  infrared  spectrum covering the wavelength   range
1.5-2.5    $\mu$m,    as    an   impressively   reddened     quasar at
$z=2.507\pm0.002$.  The mass contained within the  Einstein ring radius is
$M(r<2.1h^{-1}Kpc)=6.3\times10^{10}h^{-1}M_{\odot}$  for  $\Omega_M=1$
or  $M(r<2.4h^{-1}Kpc)=7.4\times        10^{10}h^{-1}M_{\odot}$    for
$\Omega_M=0.3$.  Our redshift measurement,  together with the recently
measured time delay (Lovell et al.), means that we are a step closer to
determining  $H_0$ from this   lens.   Converting the time delay  into
$H_0$ by  using  existing models leads   to high values of the  Hubble
parameter,           $H_0$=$65^{+15}_{-9}$ for $\Omega_M=1$ and
$H_0$=$76^{+18}_{-10}$ for $\Omega_M=0.3$.   Since the  lensing galaxy
lies very close to the center of the lensed  ring, improving the error
bars on $H_0$ will  require   not  only  a  more precise  time   delay
measurement, but also very precise astrometry of the whole system.

\end{abstract}


\section{Introduction}

The compact, flat-spectrum radio source PKS~1830-211
(\cite{sub90,jau91}) is a well-studied gravitational lens located
behind the galactic plane $(l,b)=(12,-5)$. At radio wavelengths,
PKS~1830-211 consists of two bright, flat-spectrum sources embedded in
a ring like structure with a diameter of 1\arcsec.  The radio emission
is known to be variable, and a time delay of 26 days has been
determined by monitoring the source at 8.6 Ghz (\cite{lov98}).  At
millimeter wavelengths, molecular absorption lines are detected in
both sources (\cite{wik96}, \cite{wik98}, \cite{frye97}), and are
attributed to a galaxy at $z=0.886$.  Additionally, redshifted 21cm
absorption is found at $z=0.193$ (\cite{lov96}).  The relative role of
these two systems in lensing of the radio source is unclear
(\cite{lovell98}).

The optical identification of PKS~1830-211 has long been hampered by
the presence of a nearby, bright M-star (\cite{djo92}) and by
obscuration by both the galactic plane and the lens itself.
PKS~1830-211 was finally identified in deconvolved images taken at
infrared wavelengths (\cite{cou98}).  These images show the NE
component of the source, the M-star and the SW component of the
source, which is most likely blended with the lensing galaxy.  The
source is very red: the NE component has $I-K \approx 7$, so that at K
(2.15 $\mu$m) it is much brighter than the nearby M-star, which
dominates at optical wavelengths.

In this letter, we describe how we were able to determine the redshift
of PKS~1830-211 from infrared spectroscopy carried  out with SOFI, the
new IR imager  and spectrograph  on  the ESO  New Technology Telescope
(NTT) at La Silla,  Chile (\cite{mor98}).  The  data also allow us  to
derive the extinction along  the line of  sight to PKS~1830-211 and an
estimate of the mass of the lensing galaxy at z=0.886.

\section{Observations and Reductions}

PKS~1830-211 was observed with SOFI (SOFI stands for Son OF ISAAC, the
latter is the near-IR instrument on the first 8.2m Unit Telescope of the
Very Large Telescope at Paranal, Chile) during the night of 1998 June
13. SOFI is an IR instrument capable of imaging, polarimetry, and
low resolution spectroscopy.  The detector is a Hawaii
1024$\times$1024 HgCdTe array.  Spectra were taken with the red grism,
which covers the wavelength range from 1.51 to 2.54 $\mu$m, and with
the 1\arcsec\ slit, which gives a resolution of 700 at 2.1 $\mu$m.
The slit was aligned to pass through both the NE and SW components of
source (see Fig. \ref{fig1}.).

\placefigure{fig1}

For accurate sky subtraction, the source was alternatively observed at
two positions separated by 40 arcseconds along  the slit.  At each
position, the exposure  was 180 seconds. The conditions  at the time of
the observation  were very cloudy, so only  24 minutes of  useful data
were obtained.

The night sky, which consists of bright emission lines and a thermal
continuum, is effectively removed by subtracting, from each other,
frames taken at alternative positions along the slit.  The resulting
images are then flat-fielded, registered, and combined to give a two
dimensional spectrum. The spectrum of the source is then extracted and
wavelength calibrated, which is done by fitting arc lines from a Xenon
lamp in the adaptor.  The residuals of the fit are 0.5 \AA.  The
extracted spectrum is then divided by that of a very hot star, in this
case an O6 star, to remove the strong atmospheric absorption features
that appear in infrared spectra.  In a final step, the spectrum is
multiplied with a blackbody curve that is appropriate for an O6 star.
We have used a blackbody with a temperature of 36,000~K.  At this
temperature and wavelength, the shape of the blackbody curve is not a
sensitive function of the temperature. Note that the O6 star may
contain weak Bracket and HeI lines that are introduced in the
reduction process. However, there is no evidence of them in the fully
reduced spectrum.

\section{Results}

Figure   \ref{fig2}  displays  the     reduced  spectrum     of
PKS~1830-211. Regions of significant atmospheric absorption are marked
by horizontal  lines.  Since the  atmosphere is  almost totally opaque
near 1.9 $\mu$m  and beyond 2.5$\mu$m, the data  in these regions have
been deleted.  Due to the presence  of clouds during the observations,
no absolute flux calibration could be  carried out and the spectrum in
Fig.  \ref{fig2} is shown on a relative  $F_{\lambda}$ flux scale.  We
could not use the $K$-band  flux derived by  Courbin et al.  (1998) to
put the spectrum  on an absolute flux scale  because the source is variable
in the IR.  The spectrum contains a contribution  from the M-star (see
Fig. \ref{fig1}) that lies  0\farcs5 to  the  North  of  the  NE
component of the QSO.  At 1.25 $\mu$m the M-star is twice as bright as
the NE component.  At 2.15 $\mu$m the NE component is 3 times brighter
that the M-star  (see Courbin et  al. 1998) and largely dominates  the
total flux that falls within the slit.

\placefigure{fig2}

A very strong ${\rm H}\alpha$ emission  line is seen  at a redshift of
$z=2.507  \pm 0.002$.   The dominant   source of  uncertainty in   the
redshift is due to internal instrumental flexure. Also visible, but at
lower signal to noise ratios, are {\rm H}$\beta$ and possibly the 4959
\AA\ and  5007 \AA\ lines of   [OIII].  All four  lines are  marked in
Fig.  \ref{fig2}.   The rest  equivalent width of  the ${\rm H}\alpha$
line  is 170 \AA\  and the velocity width  is 2600 \kms. Both measures
are somewhat lower than  the average value  for high  redshift quasars
(\cite{esp89}), although the range  for  high redshift quasars  are
large.   The   ${\rm H}\alpha$ line  cannot be   fitted  with a single
Gaussian.  There    are broad wings  on   either  side of   the  line,
particularly on the red side.

The  region around  the  ${\rm  H}\beta$  and  [OIII] lines is   quite
noisy. The lines may be blended with a FeII  multiplet, \#42 in Oke \&
Lauer (1979).  This multiplet is a common feature in quasar spectra at
both low and high  redshifts. However, the   strength of this  feature
varies   greatly  from    one  quasar   to    the  next (\cite{bor92};
\cite{hil93}; \cite{mur98}).   A spectrum with  better signal to noise
is required before  these  features can be positively  identified. The
region near 18,000  \AA\   is on the   edge of  the strong  absorption
feature that lies between the $H$ and $K$ IR windows.

The redshift  determined from ${\rm H}\beta$  line alone is $z=2.507$,
the  same as that determined from  ${\rm  H}\alpha$. Thus, despite the
noisy nature of the region surrounding this  line, it is unlikely that
the line at 23,000 \AA\ has been misidentified as ${\rm H}\alpha$.

The 5007 \AA\ [OIII] line is slightly blueshifted ($z=2.504$) with
respect to the ${\rm H}\alpha$ and ${\rm H}\beta$ lines.  If the
relative motion of the broad and narrow line regions is along the line
of site, the difference in the redshift corresponds to a velocity
shift of 250 km/s.  This has been observed in other quasars
(\cite{nis98}).  Other common quasar emission lines such as HeII at
4686 \AA, HeI at 5876 \AA, ${\rm H}{\gamma}$, [OIII] at 4363 \AA\ and
a FeII complex around 4500 \AA\ are not detected, although they are
within the wavelength limits of our spectrum in Figure
\ref{fig2}. However, given the strength of the ${\rm H}\beta$ line,
this is not surprising.

The  continuum  shows  a steep  increase  towards  longer wavelengths.
Between  1.6 and 2.2   $\mu$m it can  be  approximated by a power  law
$F_{\nu}  \propto \nu^{\alpha}$ with a  spectral index $\alpha \approx
-4$.  This  is considerably steeper than  the median  value of $\alpha
\approx -0.3$ for  quasar  spectra  (\cite{fra91}),  even  considering
orientation effects  (Baker  \& Hunstead,  1995),   and it implies   a
considerable amount of reddening.   Any reddening in  the O star  that
was used to remove the  atmospheric absorption features would bias the
derived reddening in PKS~1830 to  smaller and not higher values.

\section{Extinction}

By comparing the radio and near-IR flux ratios between the lensed
components of PKS~1830-211, Courbin et al.  (1998) estimated a
differential extinction in the rest frame of the lens of $E(B-V)=2.75$
between the SW and NE components.  This value is estimated from $K$
band observations and assumes that the lensing galaxy does not
contribute to the flux of the SW source (\cite{cou98}).  At the lens
redshift, the $K$ band is still in the near-IR, so we can safely
assume that extinction does not depend dramatically on galaxy
type, as it could at shorter wavelengths.  Thus, if we adopt a
galactic extinction law with $R_V=3.05$, we find that the SW component
is absorbed by $A_V=8.6$ magnitudes relative the NE component.

Our near-IR spectrum is dominated by the light from the NE component
(the SW component is 3 magnitudes fainter), and offers the opportunity
to estimate the line-of-sight absolute extinction.  For this purpose,
we do not use the slope of the continuum, which is partially
contaminated by the nearby M-star, but instead measure the Balmer
decrement, given by the line flux ratio
$F[\rm{H}_{\alpha}]/F[\rm{H}_{\beta}]$, and compare it with previous
measurements of other quasars.  Common measurements give
$F[\rm{H}_{\alpha}]/F[\rm{H}_{\beta}]\sim 4-5$ and are interpreted in
terms of moderate intrinsic reddening (e.g., \cite{hil93}) or enhanced
$\rm{H}{\alpha}$ emission due to collisional excitation
(\cite{Bak94}).  \cite{Bak95} show that the Balmer decrement depends
on the core-to-lobe flux ratio, $R$.  They find
$F[\rm{H}_{\alpha}]/F[\rm{H}_{\beta}]$ in the range 5-6 for $R >1$, and
up to 10 for $R<0.1$.  

If we assume that the $\rm{H}{\beta}$ line is uncontaminated by other
lines, $F[\rm{H}_{\alpha}]/F[\rm{H}_{\beta}]=11\pm2$. This implies
that PKS~1830-211 is significantly reddened.  If we assume further
that all the absorption occurs in the z=0.886 lens, that the Balmer
decrement of the unreddened quasar is five, and that a galactic
absorption law is applicable, we find that the NE component has
$E(B-V)\approx 1.2$ in the rest frame of the lens.  Thus, the
extinction towards the NE component in the rest frame of the lens is
$A_V\approx 3.7$ magnitudes, which implies an extinction of
$A_V\approx 12$ magnitudes towards the SW component.

We stress that these estimates are very uncertain.  They can change if
any of the following are true: (i) some fraction of the reddening is
caused by our Galaxy or by the HI absorber at $z=0.19$, (ii) the
$\rm{H}{\beta}$ line flux is overestimated, (iii) PKS~1830-211 is
reddened at the source, (iv) the SW component contributes
significantly to the flux, (v) the SW component is blended with some
other object, for example, the lensing galaxy (Courbin et al.  1998).
(vi) the galactic extinction law does not apply for the lens galaxy,
and (vii) the Balmer decrement of the unreddened quasar is not five.

\section{Discussion-Conclusions}

With the source and lens redshifts known, the mass interior to the
radio ring can be estimated from a simple point mass model.  Table 1
gives this mass for two different cosmologies. Such relatively low
masses, in the range $6-14\times 10^{10} M_{\odot}$, combined with the
fact that a very significant amount of dust is present in the
deflector, suggests that the deflector is a late type spiral.  More
detailed models (\cite{rao}), which incorporate some constraints on
the position of the lensing galaxy relative to the QSO images and on
the geometry of the source, give even lower values: $2.5-5\times
10^{10} M_{\odot}$.  The models of Kochanek \& Narayan (1992), which
use the radio structure of the lens to characterise the lensing
potential, give one dimensional velocity dispersions that range
between 150 and 210 km/s.

PKS~1830-211 is known to be variable at radio and millimeter
wavelengths (\cite{van95}, \cite{Com98}, \cite{lov98}). With the
present measurement of the redshift of the source, this quasar becomes a
good candidate for constraining $H_{0}$.  Recently, a time delay of
$26^{+4}_{-5}$ days between its two point-like components was reported
(\cite{lov98}). This is substantially smaller than the time delay of $44\pm9$
days reported by van Ommen et al. (1995).  Since the lower
estimate of the time delay was obtained from radio observations 
with high angular resolution, almost allowing for the separation of
the NE and SW components, we consider it to be more
accurate.  We use this low value with the models of \cite{rao} to
deduce $H_0$.  The derived estimates tend to be higher that those
derived from other lensed quasars, even with a large $\Omega_M$.  For
$\Omega_M=1$, $H_0$=$65^{+15}_{-9}$; for $\Omega_M=0.3$ it increases to
$H_0$=$76^{+18}_{-10}$.

The rather large uncertainty in the time delay, approximately 20\%,
translates directly into an uncertainty in $H_{0}$. This uncertainty
may be lessened by monitoring PKS~1830-211 in the near-IR.  In
particular, $K$-band observations, where the source is bright due to the
presence of redshifted H$\alpha$ emission (see Fig.  2), should
allow accurate photometry of both the NE and SW components.
Our IR observations of the quasar obtained in March and June 1998 show
that the NE and SW components have brightened by $0.^{m}35$ and
$0.^{m}25$ respectively.  Similar changes were also found at
millimeter wavelengths during the same epochs (T. Wiklind 1998,
private communication).

The strong H${\alpha}$ line can be used to search for differences
between continuum and line variability and thus can be used to
help us distinguish between intrinsic source variability and differential
magnification of the continuum versus the broad-line region due to the
passage of compact micro-lensing sources in the lensing galaxy.
Monitoring of PKS~1830-211 carried out simultaneously in $H$
(continuum) and $K$ (H$\alpha$) would provide adequate material, not
only for the measurement of $H_0$, but also for the study of
micro-lensing itself.

Newly available Near-Infrared Camera and Multiobject Spectrograph
images show that the lensing galaxy is located near to the center of
the radio ring, so that the two QSO images probe the lensing potential
at similar distances from the lens center.  We can therefore expect
that detailed modelling of the system will strongly depend on the
accuracy reached on the position of the lensing galaxy, as found in
PG~1115+080 (e.g., Keeton \& Kochanek, 1997, Courbin et al. 1997,
Impey et al. 1998).

The relative ease with  which we obtained  the redshift of this lensed
radio     source by IR  spectroscopy   lends  considerable hope to the
identification  of the other  classical radio lenses  for which source
redshifts are still outstanding.

\acknowledgments

The    authors would like  to     thank D.   Hutsem\'ekers for  useful
conversations on reddening  by the lensing  galaxy. F.C.  is supported
by  contracts ARC 94/99-178   ``Action de Recherche  Concert\'ee de la
Communaut\'e       Fran\c{c}aise''     and      P\^ole    d'Attraction
Interuniversitaire P4/05 (SSTC, Belgium).

\clearpage

\begin{table}[h]
\begin{center}
\caption{Mass interior   to the  Einstein   ring  in PKS~1830-211  for
different cosmologies.}

\vspace*{10mm}
\begin{tabular}{l c c c c}
\hline \hline 
Cosmology& $D_l D_s / D_{ls}$ & $r_E$  &  Mass \\ 
           &  ($Gpc$)           & ($Kpc$)  &   ($10^{10}M_{\odot}$) \\ \hline
$H_0=50,\Omega_M=1$ & 4.2 & 4.1 & 12.5 \\ 
$H_0=50,\Omega_M=0.3$ & 4.8 & 4.7 & 14.0   \\ 
$H_0=100,\Omega_M=1$ & 2.1 & 2.1 & 6.3   \\ 
$H_0=100,\Omega_M=0.3$ & 2.5  & 2.3 & 7.4  \\ 
\\ \hline
\end{tabular}
\tablecomments{The adopted  Einstein radius of the   radio ring is the
one  obtained from the millimeter observations  by Frye et al. (1998),
i.e.,  $r_E$=0.49\arcsec. $D_l$ and  $D_s$ denote the angular diameter
distances to the lens and   source.  $D_{ls}$ is the angular  diameter
distance between the lens and the source.}
\end{center}
\end{table}

\clearpage

\figcaption[pks1830ima.ps]{\label{fig1}  Part   of  the $K$-band  image   of
PKS~1830-211 taken with SOFI at the NTT, with the position of the slit
displayed. The seeing is  0\farcs6.  North is up and  East is left.   The
inset  shows the deconvolved  central region surrounding PKS~1830-211.
The M-star  and the NE and SW  components of PKS~1830-211  are readily
apparent in  the   deconvolved   image,  which has   a   resolution of
0\farcs2.}

\plotone{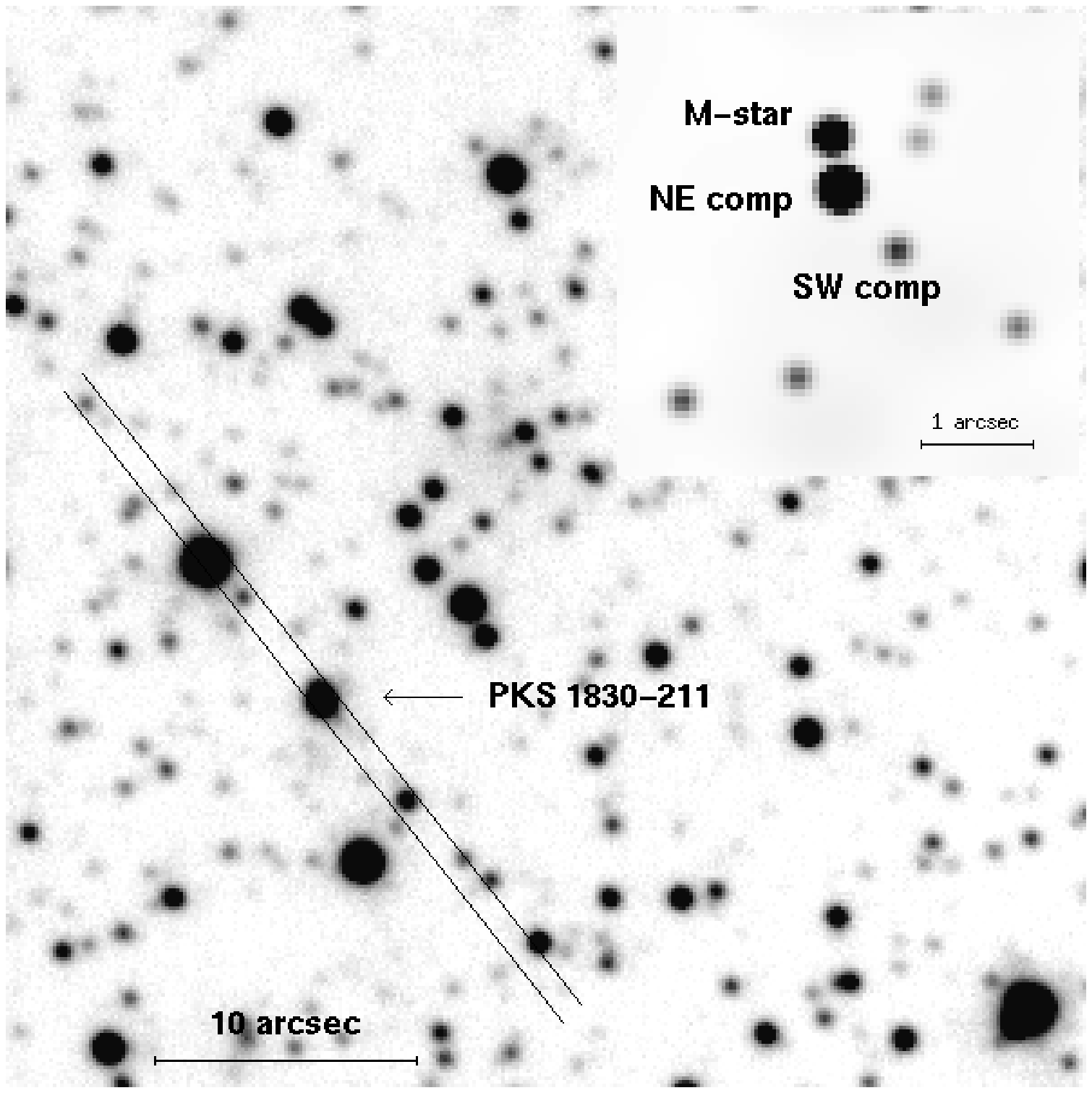}

\clearpage

\figcaption[pks1830.redfc.ps]{\label{fig2} Spectrum of PKS~1830-211.
The spectrum is plotted on a relative flux scale against observed
wavelength. Regions of strong atmospheric absorption are marked with
horizontal lines. Regions that are almost opaque are excluded.}

\plotone{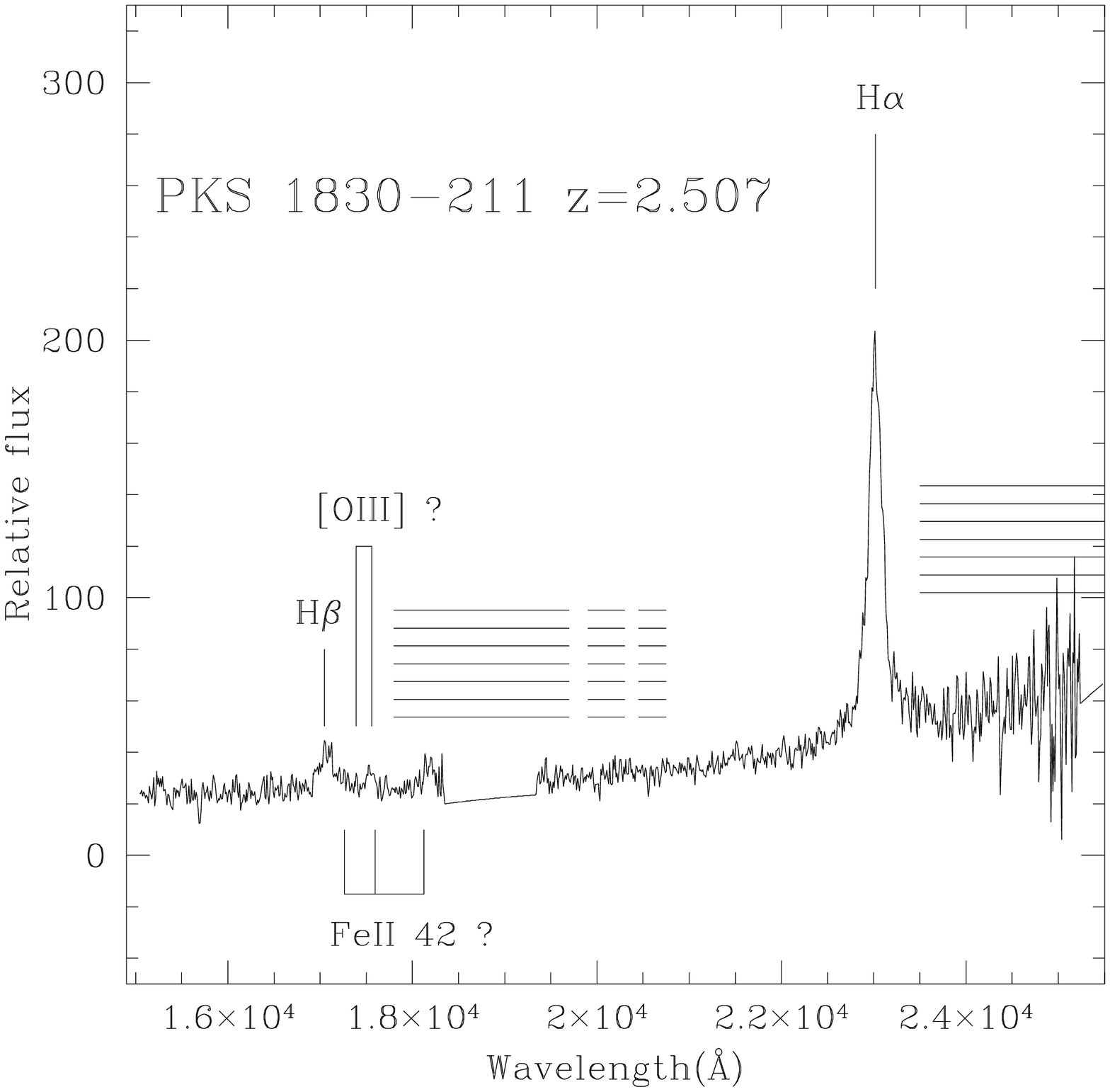}

\end{document}